\documentclass[twoside,reqno]{HERON}
\usepackage{epsfig,cite,colordvi}
\usepackage{graphicx}
\usepackage{url}
\usepackage{amssymb,amsmath,amscd,epsf}
\usepackage{times}
\usepackage{makeidx}
\usepackage[english]{babel}
\def\q{\mbox{\boldmath $q$}}

\begin{document}

\title{Relativistic Descriptions of Final-State Interactions in Quasielastic
Electron and Neutrino-Nucleus Scattering}

\runningheads{}{}

\begin{start}

\author{C. Giusti}{1}, \coauthor{A. Meucci}{1},

\index{Giusti, C.}
\index{Meucci, A.}

\address{Dipartimento di Fisica Nucleare e Teorica, 
Universit\`{a} degli Studi di Pavia and \\
INFN, Sezione di Pavia, via Bassi 6 I-27100 Pavia, Italy}{1}

\begin{Abstract}
Models developed within the framework of the relativistic impulse approximation 
for quasielastic (QE) electron and neutrino-nucleus scattering are discussed. 
Different descriptions of final-state interactions (FSI) in the 
inclusive scattering are compared: the relativistic Green's function (RGF) 
and the relativistic mean field (RMF).
The results of the models are compared with the recently measured 
double-differential charged-current QE (CCQE) neutrino MiniBooNE cross sections.
\end{Abstract}
\end{start}

\section{Introduction}
Several decades of experimental and theoretical work on electron scattering have
provided a wealth of detailed information on nuclear structure and 
dynamics~\cite{rep,book}. 
Additional information is available from neutrino-nucleus scattering.
Recently, muon
charged-current (CC) neutrino-nucleus double differential cross
sections have been measured by the MiniBooNE collaboration~\cite{miniboone}.
High-quality predictions for neutrino-nucleus cross sections
are needed for use in ongoing experimental studies
of neutrino oscillations at GeV energies and a 
proper analysis of data requires that the nuclear response to neutrino 
interactions is well under control and that the unavoidable uncertainties on 
nuclear effects are reduced as much as possible. 
%
%
Although the two situations are different, electron scattering is the best 
available guide to determine the predictive power of a nuclear model. 
Nonrelativistic and relativistic models have been used
to describe nuclear effects with different approximations. 
Relativity is however important at all 
energies, in particular at high energies, and in the energy regime of many 
neutrino experiments a fully relativistic approach is required, where not only
relativistic kinematics is considered, but also nuclear dynamics and current 
operators should be described within a relativistic framework. 

Models for the QE exclusive and inclusive electron and neutrino scattering 
are presented in this contribution. In the QE region the nuclear 
response is dominated by the mechanism of one-nucleon knockout, where the probe 
interacts with a quasifree nucleon that is emitted from the nucleus with a 
direct one-step mechanism and the remaining nucleons are spectators.
In the exclusive $(e,e'p)$ reaction the outgoing nucleon is detected in 
coincidence with the scattered electron, the residual nucleus is left in a 
specific discrete eigenstate and the final state is completely 
specified. In the inclusive $(e,e')$ scattering only the scattered
electron is detected, the final nuclear state is not determined, and the cross 
section includes all the available final nuclear states. 
The inclusive CC scattering where only the charged lepton is detected 
can be treated with the same models used for the inclusive $(e,e')$ reaction. 

For all these processes the cross section is obtained  in the one-boson 
exchange approximation from the contraction between the lepton tensor, that
under the assumption of the plane-wave approximation for the initial and the final
lepton wave functions depends  only on the lepton kinematics,  and the hadron 
tensor $W^{\mu\nu}$, whose components 
are given by bilinear products of the matrix elements of the nuclear current  
$J^{\mu}$ between the initial and final nuclear states, \ie,
\begin{equation}
W^{\mu\nu} = \sum_f \, \langle \Psi_f\mid J^{\mu}(\q) \mid \Psi_i\rangle \, 
\langle \Psi_i \mid J^{\nu\dagger}(\q)\mid \Psi_f\rangle \, 
\delta(E_i+\omega-E_f),
\label{eq.wmn}
\end{equation}
where $\omega$ and $\q$ are the energy and momentum transfer, respectively.
Different but consistent models to calculate the components of the hadron tensor 
in QE electron and neutrino-nucleus scattering are outlined in the next 
sections.

\section{The exclusive $(e,e^{\prime}p)$ reaction}

For the exclusive $(e,e^{\prime}p)$ reaction a model based on the 
distorted-wave impulse approximation (DWIA) has been developed to calculate 
the matrix elements in Eq.~(\ref{eq.wmn}). The model is based on the following 
assumptions~\cite{rep,book,bof82}: 
\newline
i) An exclusive process is considered, where the residual nucleus is left in a
discrete eigenstate $n$ of its Hamiltonian.
\newline
ii) The final nuclear state is projected onto the channel subspace spanned by 
the vectors corresponding to a nucleon, at a given position, and the residual 
nucleus in the state $n$. This assumption neglects effects of coupled channels 
and is justified by the considered asymptotic configuration of the final state.
\newline
iii) The (one-body) nuclear-current operator does not connect different channel 
subspaces and also the initial state is projected onto the selected channel 
subspace. This is the assumption of the direct-knockout mechanism 
and of the IA.
  
The amplitudes of Eq.~(\ref{eq.wmn}) are then obtained in a one-body representation as 
\begin{equation}
\lambda_n^{1/2} \langle\chi^{(-)}\mid  j^{\mu}(\q)\mid \varphi_n \rangle  \ ,
\label{eq.dko}
\end{equation}  
where $j^{\mu}$  the one-body nuclear current, $\chi^{(-)}$ is the 
single-particle (s.p.) scattering state of the emitted nucleon, $\varphi_n$ the 
overlap between the ground state of the target and the final state $n$, 
i.e., a s.p. bound state, and the spectroscopic factor $\lambda_n$ is the norm
of the overlap function, that gives the probability of removing from the target
a nucleon leaving the residual nucleus in the state $n$. 
In the model the s.p. bound and scattering states are 
eigenfunctions of a non Hermitian energy dependent Feshbach-type optical 
potential and of its Hermitian conjugate at different energies. 
In standard DWIA calculations phenomenological ingredients are usually employed: 
the scattering states are eigenfunctions of a phenomenological optical potential 
determined through a fit to elastic nucleon-nucleus scattering data and the 
s.p. bound states are obtained from mean-field potentials, or can be calculated 
in a phenomenological Woods-Saxon well.  

The model can be formulated in a similar way within nonrelativistic \cite{bof82}
DWIA and relativistic RDWIA frameworks \cite{meucci01}. 
Both the DWIA and the RDWIA have been quite successful in describing 
$(e,e^{\prime}p)$ data in a wide range of nuclei and in different
kinematics~\cite{book,meucci01,ud93,epja}.

\section{Inclusive lepton-nucleus scattering}

In the inclusive scattering where only the outgoing lepton is detected all 
elastic and inelastic channels contribute and the complex 
potential, with imaginary terms designed to reproduce just the elastic channel, 
should be dismissed. 
Different approaches have been used to account for FSI. In the approaches 
based on the RDWIA, FSI have been accounted for 
by purely real potentials. The final nucleon state has been evaluated 
with the real part of the relativistic energy-dependent optical potential (rROP), 
or with the same relativistic mean field potential considered in describing 
the initial nucleon state (RMF) \cite{Chiara03,cab1}.
However, the rROP is unsatisfactory from a 
theoretical point of view, since it is an energy-dependent potential, 
reflecting the different contribution of open inelastic channels for each 
energy, and under such conditions dispersion relations dictate that the 
potential should have a nonzero imaginary term \cite{hori}. 
On the other hand, the RMF model is based on the use of the same potential for 
bound and scattering states. 
It fulfills the dispersion relations~\cite{hori} and also the continuity 
equation. 

A different description of FSI makes use of 
Green's function (GF) techniques \cite{eenr,ee,cc,eea,eesym,acta,acta1}. 
In the GF model, under suitable approximations that are basically related to the
IA, the components of the hadron tensor are written in terms of the s.p. 
optical model Green's function. 
This result has been derived by arguments based on multiple 
scattering theory \cite{hori} or by means of projection operators techniques
within nonrelativistic ~\cite{eenr} and relativistic~\cite{cc,ee,eea} 
frameworks. The explicit calculation of the s.p. Green's function can be 
avoided\cite{eenr,ee,cc}  by its spectral representation, that is based on a 
biorthogonal expansion in terms of a non Hermitian optical potential $\cal H$ 
and of its Hermitian conjugate $\cal H^{\dagger}$. Calculations require matrix 
elements of the same type as the DWIA ones in Eq.~(\ref{eq.dko}), but involve 
eigenfunctions of both $\cal H$ and $\cal H^{\dagger}$, where the imaginary 
part gives in one case absorption and in the other case gain of flux, and 
in the sum over $n$ the total flux is redistributed and conserved.  
The GF model allows to recover the contribution of non-elastic channels starting 
from the complex optical potential that describes elastic nucleon-nucleus 
scattering data. It provides a consistent treatment of FSI in the exclusive and in 
the inclusive scattering and gives also a good description of $(e,e')$ 
data~\cite{eenr,ee,confee}. 

\begin{figure}[b]
\centering
\includegraphics[scale=0.4]{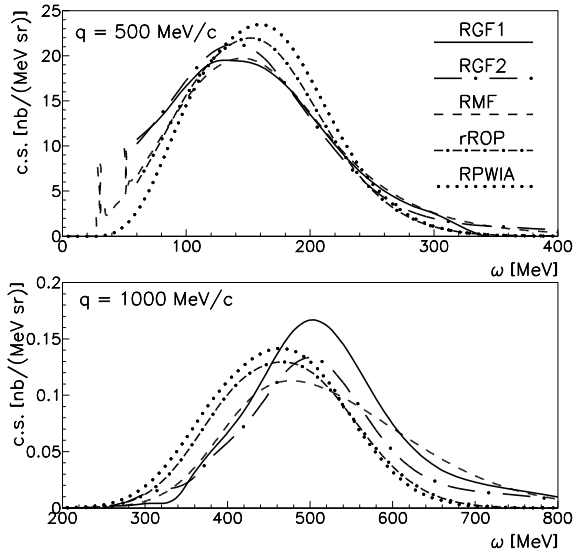}
\caption{Differential cross section of the 
$^{12}$C$(e,e^{\prime})$ reaction for an incident electron energy 
$\varepsilon = 1$ GeV and $q = $ 500 and 1000 MeV$/c$. Results for RPWIA (dotted), rROP (dot-dashed), RGF1 (solid),  
RGF2 (long dot-dashed), and RMF (dashed) are compared.} 
\label{fig1}  
\end{figure}

For the inclusive electron scattering both nonrelativistic GF~\cite{eenr,eesym} 
and relativistic RGF~\cite{ee} models have been considered, for CC 
neutrino scattering the RGF model has been adopted \cite{cc}. 
The results of the RMF and RGF models have been compared
in~\cite{confee} for the inclusive electron scattering and in~\cite{confcc} for
the inclusive CC neutrino scattering. 
An example is shown in  Figure~\ref{fig1}, where the  RGF, RMF, and rROP cross
sections of the $^{12}$C$(e,e')$ reaction calculated in a kinematics with a 
fixed value of the incident electron energy ($\varepsilon=1$ GeV), and two 
values of the momentum transfer ($q=500$ and 1000 MeV$/c$) are displayed. 
Two parameterizations of the ROP have been used for the RGF
calculations, \ie, the energy-dependent and A-dependent EDAD1 (RGF1) 
and EDAD2 (RGF2) \cite{chc}. The results of the relativistic plane-wave IA (RPWIA), 
where FSI are neglected, are also shown in the figure.

The differences between the RMF and RGF results, as well as the differences
between RGF1 and RGF2, increase with the momentum transfer.  At $q$ = 500 MeV$/c$ the three results are similar, 
both in magnitude and shape, larger differences are obtained 
at $q = 1000$ MeV$/c$. The shape of the RMF cross section shows an asymmetry, 
with a tail extending towards higher values of $\omega$, that is 
essentially due to the strong energy-independent scalar and vector potentials 
present in the RMF model.
The asymmetry towards higher $\omega$ is less significant but still visible 
for RGF1 and RGF2, whose cross sections show a
similar shape but a significant difference in magnitude. 
At $q$ = 1000 MeV$/c$ both the RGF1 and RGF2 cross sections are higher than the 
RMF one in the maximum region, but a stronger enhancement is obtained with 
RGF1, which at the peak overshoots the RMF cross section up to $40\%$ and it is 
even higher than the RPWIA result.

The differences between the  RGF1 and RGF2 results are consistent with the 
general behavior of the two ROP's and are basically due to their imaginary part. The 
real terms are very similar and the rROP cross sections are not sensitive to the 
parameterization considered. The scalar and vector 
components of the real part of the ROP get smaller with increasing energies and 
the rROP result approaches the RPWIA one for large values of $\omega$. In contrast, the imaginary part presents its 
maximum strength around 500 MeV, being also sensitive to the particular ROP 
parameterization. 

\begin{figure}[b]
\centering
\includegraphics[scale=0.4]{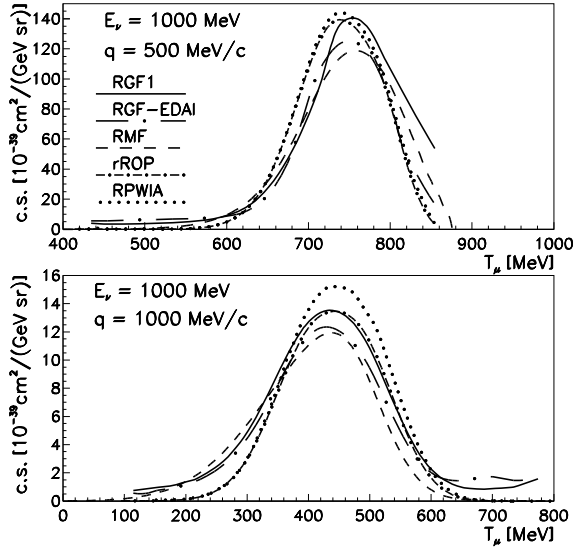}
\caption{Differential cross section of the 
$^{12}$C$(\nu_{\mu} , \mu ^-)$ reaction 
	for E$_{\nu}$ = 1000 MeV and $q$ = 500 MeV/$c$ and 1000 MeV/$c$.
 Results for RPWIA (dotted), rROP (dot-dashed) RGF1 (solid), RGF-
 EDAI (long dot-dashed), and RMF (dashed) are compared.}
\label{fig2}
\end{figure}

An example for the $^{12}$C $(\nu_{\mu},\mu^{-})$ cross section is shown in 
Figure~\ref{fig2}. For the RGF model, the RGF1 results are compared with the 
results obtained with the 
energy-dependent but A-independent EDAI potential (RGF-EDAI).
Calculations have been carried out
with the same incident lepton energy and momentum transfer as for the 
$(e,e^{\prime})$ cross sections of Figure~\ref{fig1}. 
Also in Figure~\ref{fig2} the shape of the RMF cross section 
shows an asymmetry with a tail extending towards higher values of $\omega$
(corresponding to lower values of the kinetic energy of the outgoing muon
$T_\mu$). An asymmetric shape is shown also by the RGF 
cross sections, while no visible asymmetry is given by the RPWIA and rROP 
results. Also in this case the differences between the RGF 
and rROP cross sections are consistent with the general behavior of the ROP's 
and are due to their different imaginary part.
As already shown for the $(e,e')$ reaction, the RGF yields in general a larger 
cross section than the RMF. This may reflect the influence of 
the pionic degrees of freedom, that can be included in a phenomenological way in
the imaginary part of the ROP \cite{confee,confcc}.

The results in Figure~\ref{fig2} present some differences with respect to the 
corresponding $(e,e^{\prime})$ cross sections in 
Figure~\ref{fig1}. In both cases the differences
between the results of the different models are generally larger for increasing value of 
the momentum transfer. For neutrino scattering, however, such a behavior is 
less evident and clear.  In particular, the RGF1 cross section at 
$q$ = 1000 Mev$/c$ does not show the strong enhancement in the region of the 
maximum shown in Figure~\ref{fig1}, where the RGF1 result is even larger than 
the RPWIA one. 
In the case of neutrino scattering the RGF results in the region of the 
maximum are generally larger than the RMF ones, but smaller than the RPWIA 
cross sections. 
The numerical differences between the RGF results for electron and neutrino 
scattering can mainly be ascribed to the combined effects of the weak current, 
in particular its axial term, and the imaginary part of the ROP \cite{confcc}.


\section{Scaling functions}

The comparison has been extended to the scaling properties of the different
models \cite{confee,confcc}.
Scaling ideas applied to inclusive QE electron-nucleus scattering
have been shown to work properly to high accuracy \cite{mai1,don1}.
At sufficiently high momentum transfer a scaling function is derived dividing 
the experimental $(e,e^{\prime})$ cross sections by an appropriate 
single-nucleon cross section. This is basically the idea of the IA. If this 
scaling function depends only upon one kinematical variable, the scaling 
variable, one has scaling of first kind. If the scaling function is roughly the 
same for all nuclei, one has scaling of second kind. When both kinds of scaling 
are fulfilled, one says that superscaling occurs.   
An extensive analysis of electron scattering data has shown 
that scaling of first kind is fulfilled at the left of the QE peak and broken 
at its right, whereas scaling of second kind is well satisfied at the
left of the peak and not so badly violated at its right. A phenomenological 
scaling function $f_L^{exp}(\psi')$ has been extracted from data of the 
longitudinal response in the QE region. The dimensioneless scaling
variable $\psi'(q,\omega)$ is extracted from the relativistic Fermi gas (RFG) 
analysis that incorporates the typical momentum scale for the selected
nucleus \cite{mai1,cab1}.
Although many models based on the IA
exhibit superscaling, only a few
of them are able to reproduce the asymmetric shape of $f_L^{exp}(\psi')$ with a significant
tail extended to high values of $\omega$ (large positive values of $\psi'$).
One of these is the RMF model. In contrast, the RPWIA and the rROP lead to 
symmetrical-shape scaling 
functions which are not in accordance with data analysis \cite{cab1,cab2}.
The scaling function of the RGF and RMF are very similar for lower values of 
the momentum transfer ($q=500-700$ MeV/$c$) and in good agreement with the
phenomenological function \cite{confee}, the asymmetric tail of the data and 
the strength at the peak are fairly reproduced by both models, while visible
discrepancies appear increasing $q$. 

\begin{figure}[b]
\centering
\includegraphics[scale=0.4]{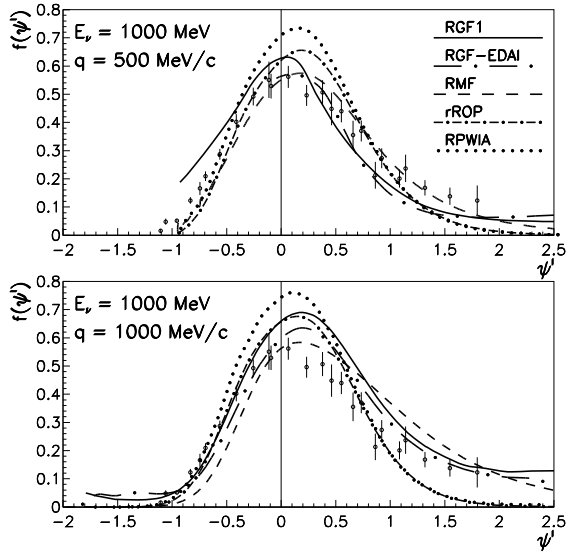}
\caption{Scaling function of the $^{12}$C$(\nu_{\mu} , \mu ^-)$ reaction 
	for incident neutrino 
	energy E$_{\nu}$ = 1000 MeV and $q$ = 500 MeV/$c$  and 
	1000 MeV/$c$. 
Line convention as in Figure~\ref{fig2}}
\label{fig3}
\end{figure}

In Figure~\ref{fig3} the scaling function extracted from the differential cross 
sections of the $^{12}$C$(\nu_{\mu} , \mu ^-)$ reaction evaluated for 
different models at $q=500$ and 1000 MeV/$c$ (shown in Figure~\ref{fig2}) are
compared with the phenomenological scaling function extracted from the analysis of 
$(e,e^{\prime})$ data.
Both the RMF and the RGF models give an asymmetric shape, with a tail in the 
region of positive $\psi^{\prime}$. In contrast, the RPWIA and the rROP results 
do not show any significant asymmetry.
As a general remark, these results for the scaling functions follow similar
trends to those already applied to the behavior of the cross sections
in Figure~\ref{fig2}.

\begin{figure}[b]
\centering
\includegraphics[scale=0.45]{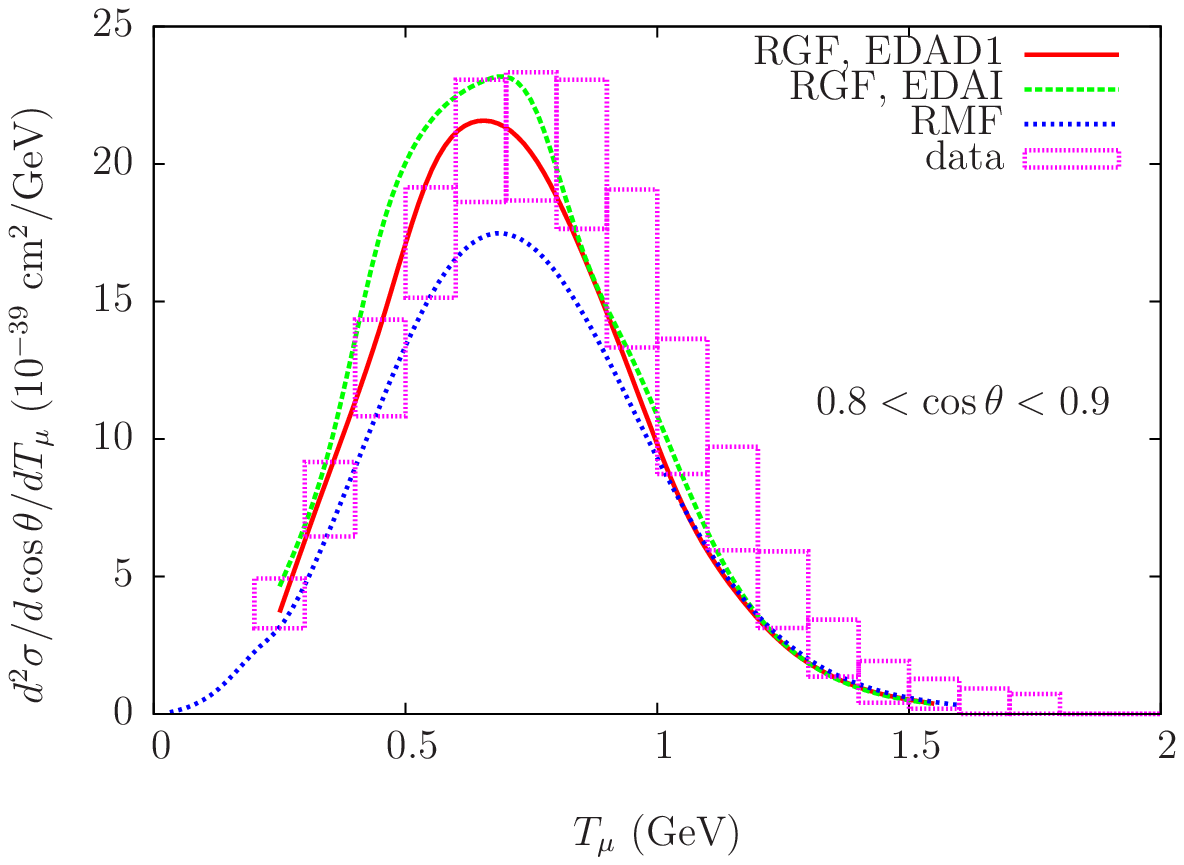}
\includegraphics[scale=0.45]{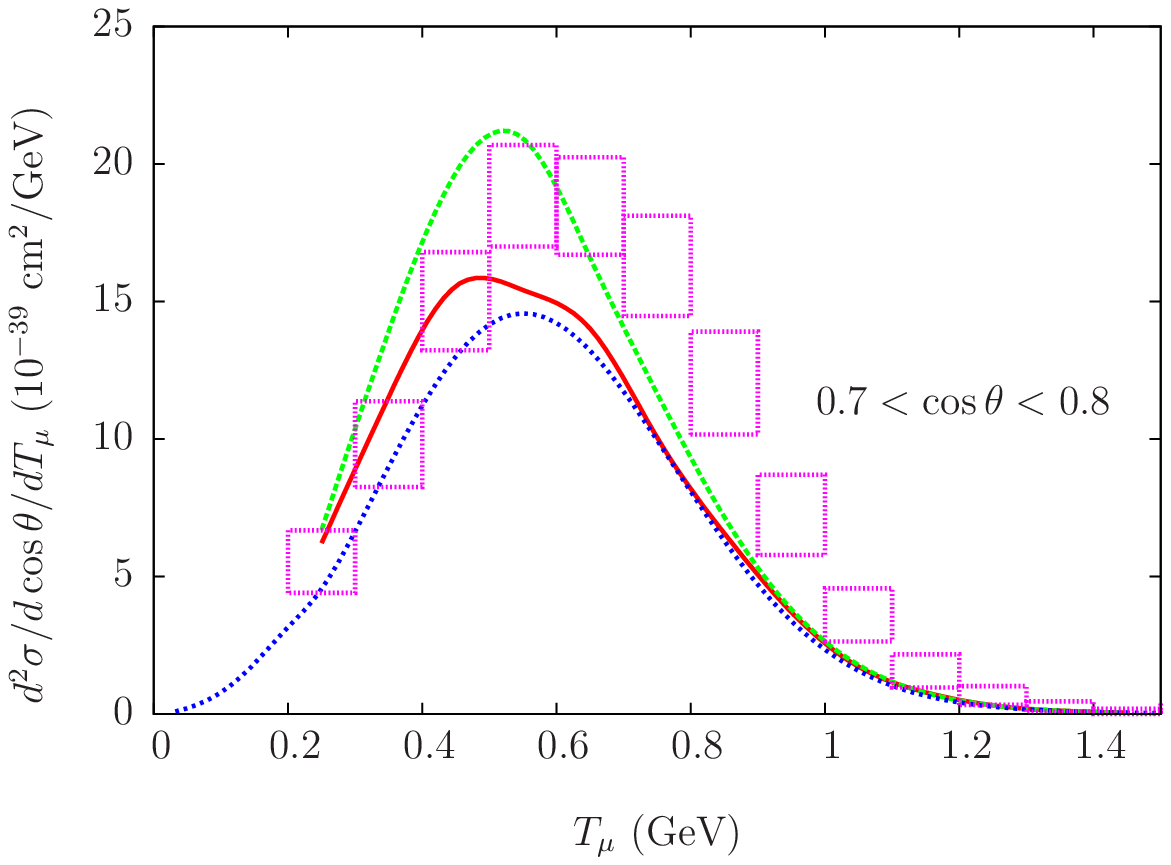}
\includegraphics[scale=0.45]{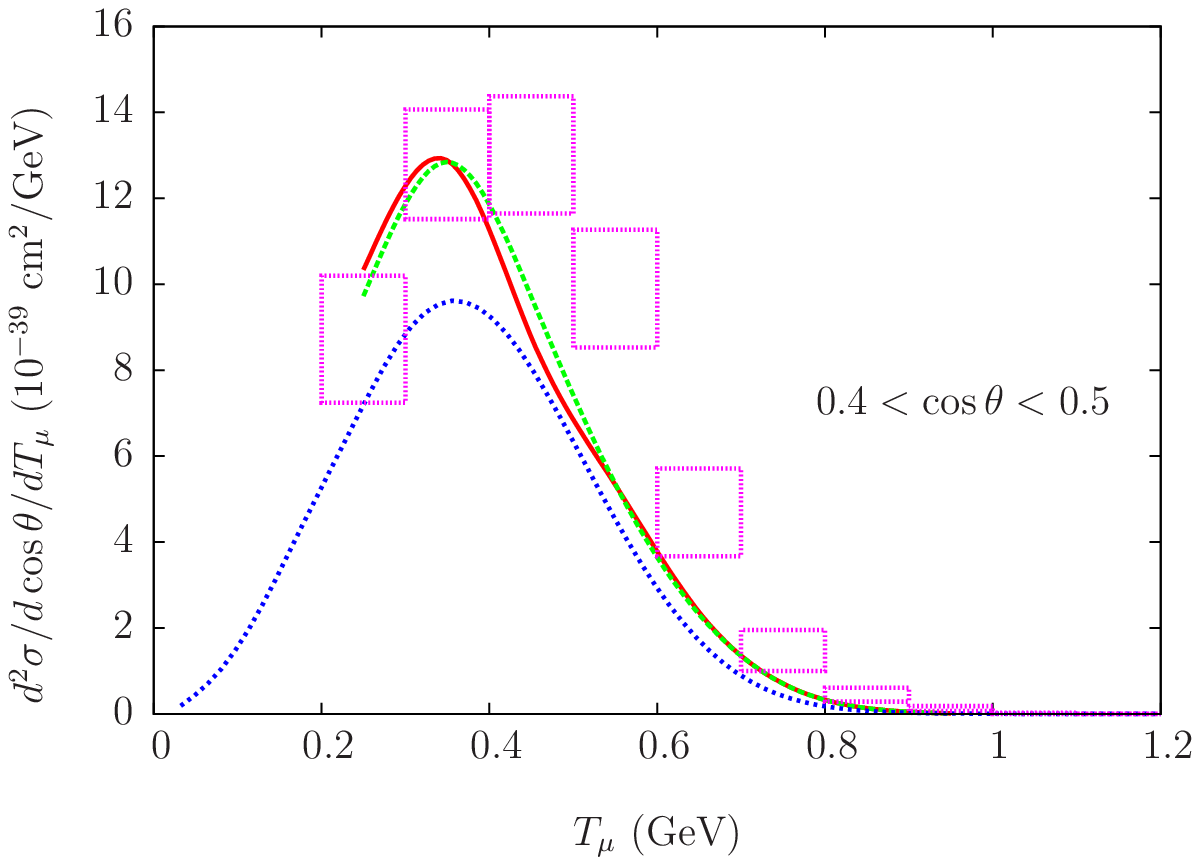}
\caption{Flux-averaged double-differential cross section per 
target nucleon for the CCQE $^{12}$C$(\nu_{\mu} , \mu ^-)$ reaction calculated 
in the RMF (blue line), the RGF1 (red), and 
	RGF-EDAI (green) and displayed versus $T_\mu$ for various bins 
	of $\cos\theta$. In all the calculations the standard value of the 
	nucleon axial mass, {\it i.e.}, $M_A = 1.03$ GeV/$c^2$ has been used.
The data are from MiniBooNE \cite{miniboone}.
}
\label{fig4}
\end{figure}
\begin{figure}[b]
\centering
\includegraphics[scale=0.6]{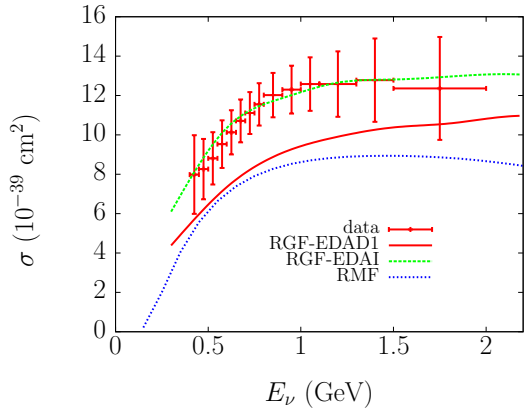}
\caption{Total CCQE cross section per neutron versus the neutrino energy. 
The cross sections calculated in the RMF (blue line), RGF EDAD1 (red), and 
RGF EDAI (green) potentials are compared with the flux unfolded MiniBooNE 
data of \cite{miniboone}.}
\label{f5}
\end{figure}

\section{Comparison with Charged-Current Quasielastic MiniBooNE data}

The CCQE $^{12}$C$(\nu_{\mu} , \mu ^-)$ cross sections 
recently measured by the MiniBooNE collaboration \cite{miniboone} have raised 
debate over the role of the theoretical ingredients entering the description of 
the reaction. The experimental cross section is underestimated by the RFG 
model, and by other more sophisticated models, unless the nucleon axial mass 
$M_A$ is significantly enlarged (1.35 GeV/$c^2$ in the RFG) with respect to the 
accepted world average value (1.03 GeV/$c^2$ \cite{Bern02,bodek}).
Before drawing conclusions about the value of the axial mass it is 
however important to evaluate carefully the contributions of all the nuclear
effects. 
Within the QE kinematic domain the treatment of FSI is essential for
the comparison with data. 
The comparison between the results of the RMF and RGF 
models can be helpful for a deeper understanding of the role played by FSI in 
the analysis of CCQE data \cite{compmini}.

The CCQE double-differential $^{12}$C $(\nu_{\mu},\mu^{-})$ cross sections 
averaged over the neutrino flux as a function of $T_\mu$ for various bins 
of $\cos\theta$, where $\theta$ is the muon scattering angle, are shown in 
Figure~\ref{fig4}.
The RMF results yield reasonable agreement with data for small angles and low muon energies, 
the discrepancy becoming larger as $\theta$ and $T_\mu$ increase. The shape 
followed by the RMF and RGF cross sections fits well the slope shown by the 
data. The two models yield close predictions at
larger values of $T_\mu$, for all the bins of $\cos\theta$ shown in 
the figure. The RGF cross sections are generally larger than the RMF ones. 
The differences increase approaching the peak region, where the additional
strength shown by the RGF produces cross sections in reasonable agreement 
with the data. The differences between the RGF-EDAI and RGF-EDAD1 results 
are enhanced in the peak region and are in general 
of the order of the experimental errors.

In Figure~\ref{f5} the total CCQE cross section per neutron obtained in 
the RMF and RGF models are displayed as a function of the neutrino energy 
and compared with the ``unfolded'' experimental data~\cite{miniboone}. 
The RMF results underpredict the MiniBoone cross section. Larger 
cross sections, in particular for larger values of $E_\nu$, are obtained in the 
RGF with both optical potentials. Visible differences are obtained between 
the RGF-EDAI and the RGF-EDAD1 results, being RGF-EDAI in good agreement with 
the shape and magnitude of the experimental cross section and RGF-EDAD1 above 
RMF but clearly below the data. These differences are due to 
the different imaginary parts of the two ROP's, 
particularly for the energies considered in kinematics with the lowest 
$\theta$ and the largest $T_\mu$. These kinematics, which were not considered 
in previous calculations \cite{cc,confcc}, give large contributions to the 
total cross section and emphasize the differences between the RGF predictions 
with the two optical potentials. 
We notice that EDAI is a single-nucleus parameterization, which does have an edge 
in terms of better reproduction of the elastic proton-$^{12}C$ phenomenology 
\cite{chc} compared to EDAD1, and also leads to CCQE results in better 
agreement with data.

The RMF model generally underpredicts the data \cite{amaro11b}. In contrast, 
the RGF can give cross sections of the same magnitude as the experimental ones 
without the need to increase the standard value of the axial mass. 
The larger 
cross sections arise in the RGF model from the translation to the inclusive 
strength of the overall effect of inelastic channels.

These results confirm that before drawing conclusions about the comparison 
with CCQE MiniBoone data and the need to increase the axial mass, the relevance  
of all nuclear effects must be investigated. 
A careful evaluations of the relevance of multi-nucleon 
emission \cite{Martini,Nieves} and of some non-nucleonic 
contributions \cite{Tina10} would be helpful to clarify the content of the
enhancement of the CCQE cross sections obtained in the RGF model.
A better determination of a phenomenological ROP 
which closely fullfills the dispersion relations deserves further investigation.

\section*{Acknowledgements}
We thank F.D. Pacati, F. Capuzzi, J.A. Caballero, J.M. Ud\'{\i}as, and M.B. 
Barbaro for the fruitful collaborations that led to the results reported in 
this contribution.


\begin{thebibliography}{99}

\bibitem{rep}
S. Boffi, C. Giusti, and F.D. Pacati, \textit{Phys. Rep.} \textbf{226} (1993) 
1-101.
 
\bibitem{book}
S. Boffi, C. Giusti, F.D. Pacati, and M. Radici \textit{Electromagnetic Response 
of Atomic Nuclei} Oxford Studies in Nuclear Physics, Vol. 20, Clarendon Press,
Oxford (1996).

\bibitem{miniboone}
A.A. Aguilar-Arevalo \etal,
\textit{Phys. Rev.D} \textbf{81} (2010) 092005-1-22. 

\bibitem{bof82}
S. Boffi \etal,
\textit{Nucl. Phys. A} \textbf{379} (1982) 509-522.
 
\bibitem{meucci01}
A. Meucci, C. Giusti, and F.D. Pacati, \textit{Phys. Rev.C} \textbf{64} (2001) 
014604-1-10.
 
\bibitem{ud93}
J.M. Ud\'{\i}as \etal,
\textit{Phys. Rev. C} \textbf{48} (1993) 2731-2739.

\bibitem{epja}
M. Radici, A. Meucci, and W.H. Dickhoff, \textit{Eur. Phys. J. A} \textbf{17}
(2003) 65-69. 

\bibitem{Chiara03}
C. Maieron \etal,
\textit{Phys. Rev. C} \textbf {68} (2003) 048501-1-4.

\bibitem{cab1}
J.A. Caballero, \textit{Phys. Rev. C} \textbf{74} (2006) 015502-1-12.

\bibitem{hori}
Y. Horikawa, F. Lenz, and N.C. Mukhopadhyay, \textit{Phys. Rev. C} \textbf{22}
(1980) 1680-1695.

\bibitem{eenr}  
F. Capuzzi, C. Giusti, and F.D. Pacati \textit{Nucl. Phys.A} \textbf{524} 
(1991) 681-705

\bibitem{ee} 
A. Meucci \etal, 
\textit{Phys. Rev.C} 
\textbf{ 67} (2003) 054601-1-12.

\bibitem{cc}
A. Meucci, C. Giusti, and F.D. Pacati \textit{Nucl. Phys. A} \textbf{739} 
(2004) 277-290.
 
\bibitem{eea}
A. Meucci, C. Giusti, and F.D. Pacati \textit{Nucl. Phys. A} \textbf{756} (2005)
359-381.
 
\bibitem{eesym} 
F. Capuzzi \etal, 
\textit{Ann. Phys.} \textbf{317} (2005) 492-529.

\bibitem{acta}
A. Meucci, C. Giusti, and F.D. Pacati, \textit{Acta Phys. Polon. B} 
\textbf{37} (2006) 2279-2286.

\bibitem{acta1}
A. Meucci, C. Giusti, and F.D. Pacati, \textit{Acta Phys. Polon. B} \textbf{40} 
(2009) 2579-2584.


\bibitem{confee}
A. Meucci \etal,
\textit{Phys. Rev. C} \textbf {80} (2009) 024605-1-12.

\bibitem{chc} 
E.D. Cooper \etal, 
\textit{Phys. Rev.C }
\textbf{47} (1993) 297-311.

\bibitem{confcc}
A. Meucci \etal,
\textit{Phys. Rev. C} \textbf {83} (2011) 064614-1-10.

\bibitem{cab}
J.A. Caballero \etal,
\textit{Phys. Rev. Lett.} \textbf {95} (2005) 252502-1-4.

\bibitem{mai1}
C. Maieron, T.W. Donnelly, and I. Sick,
\textit{Phys. Rev. C} \textbf{65} (2002) 025502-1-15.

\bibitem{don1}
T.W. Donnelly and I. Sick,
\textit{Phys. Rev. Lett.} \textbf{82} (1999) 3212-3215;
\textit{Phys. Rev. C} \textbf{60} (1999) 065502-1-16.


\bibitem{cab2}
J.A. Caballero \etal,
\textit{Phys. Rev. Lett.} \textbf{95} (2005) 252502-1-4.

\bibitem{Bern02}
V. Bernard, L. Elouadrhiri and U.G. Meissner, 
\textit{J. Phys.G} {\bf 28} (2002) R1-R35.

\bibitem{bodek}
A. Bodek \etal,
\textit{Eur. Phys. J. C} \textbf{53} (2008) 349-354.

\bibitem{compmini}
A. Meucci \etal, 
\textit{Preprint arXiv:1107.5145 [nucl-th]} (2011).

\bibitem{amaro11b}
J.E. Amaro \etal, 
\textit{Phys.Rev. D} \textbf{84} (2011) 033004-1-8.

\bibitem{Martini}
M. Martini \etal,
\textit{Phys. Rev. C} \textbf{80} (2009) 065501-1-16;
\textit{Phys. Rev. C} \textbf{81} (2010) 045502-1-5.

\bibitem{Nieves}
 J. Nieves, I. Ruiz Simo,  and M.J. Vicente Vacas,
\textit{Phys. Rev. C} \textbf{83} (2011) 045501-1-19;
\textit{Preprint arXiv:1106.5374 [nucl-th]} (2011).

\bibitem{Tina10}
T. Leitner and U. Mosel,
\textit{Phys. Rev. C} \textbf{81} (2010) 064614-1-10.

\end{thebibliography}
\end{document}